\documentclass[12pt]{article}          %%
\usepackage{graphics}
\usepackage{graphicx}
\input amssym.tex

\usepackage{%
pstricks,%
pst-plot,%
epsf%
}
\usepackage{cite}
\usepackage{graphicx}
\usepackage{array,dcolumn}      % belongs to LaTeX2e tools
\usepackage{axodraw}

\newcommand{\hs}{\hspace*{0.5cm}}

\newcommand{\be}{\begin{equation}}
\newcommand{\ee}{\end{equation}}
\newcommand{\bea}{\begin{eqnarray}}
\newcommand{\eea}{\end{eqnarray}}
\newcommand{\bary}{\begin{array}}
\newcommand{\eary}{\end{array}}
\newcommand{\bit}{\begin{itemize}}
\newcommand{\eit}{\end{itemize}}
\newcommand{\ben}{\begin{enumerate}}
\newcommand{\een}{\end{enumerate}}
\newcommand{\crn}{\nonumber \\}

\newcommand{\al}{\alpha}

\newcommand{\bet}{\beta}
\newcommand{\ga}{\gamma}

\newcommand{\fr}{\frac}

\newcommand{\bc}{\begin{center}}
\newcommand{\ec}{\end{center}}

\newcommand{\ep}{\epsilon}

\newcommand{\La}{\Lambda}

\newcommand{\si}{\sigma}

\newcommand{\Om}{\Omega}

\begin{document}

\bc {\Large \bf RADION PRODUCTION IN $\gamma e^-$ COLLISIONS}

\vspace*{1cm}

{\bf D.V.Soa$^{a,}$\footnote{dvsoa@hnue.edu.vn}, D.T.L.Thuy$^{a,}$
N.H.Thao$^{b,}$
 and  T.D.Tham$^{b,}$\\}

\vspace*{0.5cm}

$^a$ Department of Physics, Hanoi University of Education, Hanoi, Vietnam \\
$^b$ Institute of Physics, VAST, P. O. Box 429, Bo Ho, Hanoi
10000, Vietnam \\

\ec
\begin{abstract}
We analyze the potential of Compact Linear Colliders (CLIC) based
on the $\gamma e^-$ collisions to search for the radion in the
Randall - Sundrum (RS) model, where compactification radius of the
extra dimension  is stabilized by the radion, which is a scalar
field lighter than the graviton Kaluza-Klein states. The  radion
production in the high energy $\gamma e^-$ colliders
 with the polarization of the electron beams are calculated in detail.
 Numerical evaluation shows that if the radion mass is not too heavy with the mass order
 of $GeV$ then the reaction can give observable cross section in future colliders at the
 high degree of polarization.
\end{abstract}


PACS number(s): 12.60.Cn, 12.60.Fr, 12.10.Kt, 14.70.Br\\

\noindent

\section{Introduction}
There has been a lot attention devoted to models of physics above
the weak scale utilizing extradimensions in solving the hierarchy
problem. Recently the scenario proposed by Randall and Sundrum
(RS) \cite{rs} can solve the hierarchy problem by localizing all
the standard model(SM) particles on the IR brane.

The RS model predicts a Kaluza-Klein tower of gravitons and
graviscalar, called radion, which stabilize the size of the extra
dimension without fine tuning of parameters and is the lowest
gravitational excitation in this scenario. The motivation for
studying the radion is twofold. Firstly, the radion may turn out
to be the lightest new particle in the RS-type setup, which
implies that the radion detection in experiments will be the first
important signature of the RS model. In addition, the
phenomenological similarity and potential mixing of the radion and
Higgs boson warrant detailed study in order to facilitate
distinction between the radion and Higgs signals at colliders.
Much research has been done on understanding possible mechanisms
for radius stabilization and the phenomenology of the
radion field\cite{csaba1,rradion,rradion1,tana,domi,sae,csabm,rhiggs1,csaki}.\\
In Refs. \cite{cheun,cheun1}, authors have considered the
associated production of radion with Higgs boson at $e^-e^+$  and
$\gamma\gamma$ colliders. Recently several authors have also
discussed the search of radion in inclusive processes at Tevaron
and LHC \cite{hoom,gon}. In our previous works \cite{stl,sblk}, we
have calculated the production cross-sections of new gauge bosons
in the 3-3-1 models in the high energy collisions.

In this paper we will consider the radion production in the high
energy $\gamma e^-$ collisions with the polarization of the
electron beams. The polarization of electron (or positron) beams
at the colliders gives a very effective means to control the
effect of the SM processes on the experimental analyses. Beam
polarization is also an indispensable tool in identifying and
studying new particles
and their interactions.\\
The organization of this paper is follows. In Sec. \ref{rsmodel}
we give a review of the RS model. In Sec. \ref{rphph} we present
the coupling of radion to photons. Section \ref{colli} is devoted
to the radion production in $\gamma e^-$ colliders. Finally, we
summarize our results and make conclusions in the last
section--Sec. \ref{culc}.

\section{\label{rsmodel}A review of RS  model} The RS model is
based on a 5D spacetime with non-factorizable geometry \cite{rs}.
The single extradimension is compactified on a $S^1/Z_2$ orbifold
of which two fixed points accommodate two three-branes (4D
hyper-surfaces), the Planck brane at $y= 0$ and TeV brane at $y =
1/2$. The ordinary 4D Poincare invariance is shown to be
maintained by the following classical solution to the Einstein
equation \be ds^2 = e^{-2\si(y)}\eta_{\mu \nu} dx^\mu dx^\nu -
b_0^2 d y^2, \hs \si(y) = m_0 b_0 |y|, \label{rsdt1} \ee where
$x^\mu$ $(\mu=0,1,2,3)$ denote the coordinates on the 4D
hyper-surfaces of constant $y$ with metric $\eta_{\mu \nu}=
\mathrm{diag}(1,-1,-1,-1) $. The $m_0$ and $b_0$ are the
fundamental mass parameter and compactification radius,
respectively.
 Gravitational fluctuations about the RS metric,
\be \eta_{\mu \nu} \rightarrow g_{\mu \nu} = \eta_{\mu \nu} + \ep
h_{\mu \nu}(x,y), \hs b_0 \rightarrow b_0 + b(x),\hs \ \ee yield
two kinds of new phenomenological ingredients on the TeV brane:
the KK graviton modes $h^{(n)}_{\mu \nu}(x)$ and the canonically
normalized radion field $\phi_0(x)$, respectively defined as \be
h_{\mu \nu}(x,y) =\sum_{n=0}^\infty h^{(n)}_{\mu \nu}(x)
\fr{\chi^{(n)}(y)}{\sqrt{b_0}},\hs
\phi_0(x)=\sqrt{6}M_{\mathrm{Pl}}\Om_b(x), \ee
 where $\Om_b(x) \equiv e^{-m_0[b_0 +b(x)]/2}$.
 The 5D Planck mass $M_5\ (\ep^2 = 16 \pi G_5 = 1/M^3_{5})$ is
 related to its 4D one ($M_{\mathrm{Pl}} \equiv 1/\sqrt{8 \pi G_\mathrm{N}}$) by
\be \fr{M^2_{\mathrm{Pl}}}{2} = \fr{1 - \Om^2_0 }{\ep^2 m_0}. \ee
Here $\Om_0 \equiv  e^{-m_0 b_0/2}$ is known  as the warp factor.
Because our TeV brane is arranged to be at $y = 1/2$, a
canonically normalized scalar field has the mass multiplied by the
warp factor, i.e, $m_{\mathrm{phys}}=\Om_0 m_0 $.
 Since the moderate value of  $m_0b_0/2 \simeq 35$ can generate TeV
 scale physical mass, the gauge hierarchy problem is explained.

The 4D effective Lagrangian is then \be \mathcal{L} =
-\fr{\phi_0}{\La_\phi}T^\mu_\mu -\fr{1}{\hat{\La }_W} T^{\mu
\nu}(x)\sum_{n=1}^\infty h^{(n)}_{\mu \nu}(x), \ee  where
$\La_\phi\equiv\sqrt{6}M_{\mathrm{Pl}}\Om_0$ is the VEV of the
radion field, and $\hat{\La}_W\equiv\sqrt{2}M_{\mathrm{Pl}}\Om_0$.
The $T^{\mu \nu}$ is the energy-momentum tensor of the TeV brane
localized SM fields. The $T^\mu_\mu$ is the trace of the
energy-momentum tensor, which is given at the tree level
as~\cite{sae,csabm} \be T^\mu_\mu = \sum_{f} m_f \bar{f}f - 2
m^2_W W^{+}_{\mu}W^{-\mu} - m^2_Z Z_{\mu}Z^{\mu}+(2m^2_{h_0} h_0^2
-\partial_{\mu}h_0 \partial^{\mu}h_0) + \cdots \ee

The gravity-scalar  mixing arises at the  TeV-brane
by~\cite{rhiggs1} \be S_\xi = -\xi \int d^4 x
\sqrt{-g_{\mathrm{vis}}} R(g_{\mathrm{vis}}) \hat{H}^\dagger
\hat{H}, \ee where $R(g_{\mathrm{vis}})$ is the Ricci scalar for
the induced metric on the visible brane or TeV brane,
$g_{\mathrm{vis}}^{\mu \nu} = \Om^2_b(x)(\eta^{\mu \nu} + \ep
h^{\mu \nu})$. $\hat{H}$ is the Higgs field before re-scaling,
i.e., $ H_0 = \Om_0 \hat{H}$. The parameter $\xi$ denotes the size
of the mixing term. With $\xi \neq 0$, there is neither a pure
Higgs boson nor pure radion mass eigenstate. This $\xi$ term mixes
the $h_0$ and $\phi_0$ fields into the mass eigenstates $h$ and
$\phi$ as given by~\cite{chi,rhiggs1} \bea
\left(%
\begin{array}{c}
  h_0 \\
  \phi_0\\
\end{array}%
\right) & = &
\left(%
\begin{array}{cc}
 1 & -6\xi \ga/Z \\
 0 & 1/Z \\
\end{array}%
\right)
\left(%
\begin{array}{cc}
\cos \theta & \sin \theta \\
 -\sin \theta &  \cos \theta \\
\end{array}%
\right) \left(%
\begin{array}{c}
 h \\
 \phi\\
\end{array}%
\right)=
\left(%
\begin{array}{cc}
   d &  c \\
 b &  a \\
\end{array}%
\right) \left(%
\begin{array}{c}
   h \\
  \phi\\
\end{array}%
\right), \label{rsd5a}
 \eea where
 \bea
\ga &\equiv& v_0/\La_\phi,\hs Z^2  \equiv  1 - 6 \xi \ga^2 (1+6
\xi) = \bet - 36 \xi^2 \ga^2,\hs
 \bet \equiv 1 - 6 \xi \ga^2,\crn
 a&\equiv&\fr{\cos \theta}{Z}, \hs
b\equiv -\fr{\sin \theta}{Z},\hs c\equiv \sin \theta - \fr{6\xi
\ga}{Z}\cos \theta,\hs d \equiv\cos \theta +
 \fr{6\xi \ga}{Z}\sin \theta. \label{rsd2}\eea The mixing angle $\theta$ is defined by
\bea \tan 2 \theta &=& 12 \ga \xi Z \fr{m^2_{h_0}}{m^2_{h_0}(Z^2
-36 \xi^2 \ga^2)-m^2_{\phi_0}}. \label{rsd3} \eea The new fields
$h$ and $\phi$ are mass eigenstates with masses \be m^2_{h,\phi} =
\fr{1}{2 Z^2}\left[m^2_{\phi_0} + \bet m^2_{h_0} \pm
\sqrt{(m^2_{\phi_0}+ \bet m^2_{h_0} )^2 - 4 Z^2
m^2_{\phi_0}m^2_{h_0} }\right]. \label{rsd4} \ee

The mixing between the states enable decays of the heavier
eigenstate into the lighter eigenstates if kinematically
 allowed. Overall, the production cross-sections, widths and relative branching
 fractions can all be affected
 significantly by the value of the mixing parameter $\xi$~\cite{rhiggs1, csabm, cheun}.
 There are also two algebraic constraints on the value of $\xi$.
 One comes from the requirement that
the roots of the inverse functions of Eq.(\ref{rsd4}) are
definitely  positive. Suggesting that the Higgs boson is heavier,
we get \bea \frac{m^2_{h}}{m^2_{\phi}} > 1+
\frac{2\bet}{Z^2}\left(1- \frac{Z^2}{\bet}\right) +
\frac{2\bet}{Z^2} \left[1-
\frac{Z^2}{\bet}\right]^{1/2}\label{rsd41}.\eea The other one is
from the fact that the
 $Z^2$ is the coefficient of the radion kinetic term after undoing the
 kinetic mixing. It is therefore required to be positive ($Z^2 > 0$)
 in order to keep the radion kinetic term  definitely  positive, i.e. \bea
-\frac{1}{12}\left(1+\sqrt{1+\frac{4}{\ga^2}}\right)
 < \xi < \frac{1}{12}\left(\sqrt{1+\frac{4}{\ga^2}}-1\right) \label{rsd42} . \eea

We now discuss the previous estimations on some model parameters.
All phenomenological signatures of the RS model including the
radion - Higgs mixing are specified by five parameters \be
\La_\phi , \ \fr{m_0}{M_{\mathrm{Pl}}}, \ m_h , \ m_\phi , \ \xi.
\ee For  the reliability of the  RS  solution, the ratio $\
\fr{m_0}{M_{\mathrm{Pl}}}$ is usually taken around $0.01\leq\
\fr{m_0}{M_{\mathrm{Pl}}}\leq 0.1$ to avoid too large
 bulk curvature ~\cite{davou}. Therefore, we
consider the case of  $\Lambda_\phi = 5\ \textrm{TeV}$ and $\
\fr{m_0}{M_{\mathrm{Pl}}}= 0.1$, where the effect of radion on the
oblique parameters is small~\cite{cskim}. In the following, let us
choose $\xi = 0, \pm1/6$, in agrement with those in
Ref.~\cite{rhiggs1} with $\xi \ga\ll 1, Z^2\approx 1$.

\section{\label{rphph}Radion coupling to photons}
For the massless gauge bosons such as photon and gluon, there are
no large couplings to the radion because there are no
brane-localized mass terms. However, the potentially large
contributions to these couplings may come from the loop effects of
the gauge bosons, the higgs field and the top quark as well as the
localized trace anomalies. We lay out the necessary radion-photon
coupling \bea
 \mathcal{L}_{\ga \ga \phi}& = & \fr 1 2  c_{\phi \ga \ga} \phi F_{\mu \nu}F^{\mu \nu}, \label{rsd6}
\eea with \be c_{\phi \ga \ga} =  -\fr{\al}{4 \pi
\La_\phi}\left\{a(b_2+ b_Y) -  a_{12}[F_1(\tau_W) +
 4/3 F_{1/2}(\tau_t)]\right\},\ee
 where $b_2= 19/6, b_Y = -41/6$ are the $\mathrm{SU}(2)_L\otimes
 \mathrm{U}(1)_Y$ $\beta$-function coefficients in the SM, and
 $a_{12}= a + c/\gamma$, $\tau_{t}=4m^2_{t}/q^2$ and $\tau_{W}=4m^2_{W}/q^2$.

The form factors $F_{1/2}(\tau_t)$ and $F_1(\tau_W)$ are given by
\bea F_{1/2}(\tau) & =& -2 \tau[1+(1-\tau) f(\tau)],\crn F_1(\tau)
&=& 2 + 3\tau+3\tau(2-\tau) f(\tau), \label{rsd7} \eea with \bea
f(\tau) = \left\{
\begin{array}{l}
\arcsin^2(1/\sqrt{\tau}),\hs \hs \hs \hs \hs  \tau \geq 1 ,
\\
-\fr 1 4 \left[\ln\left( \fr{1+\sqrt{1-\tau}}{1-\sqrt{1-\tau}}
\right) - i \pi \right]^2  \hs \hs \! \tau < 1  .
\end{array}
\right. \eea The important property of $ F_{1/2}(\tau)$ is that,
for $\tau > 1$, it very quickly saturates to $-4/3$, and to $0$
for $\tau < 1$.  $F_1(\tau)$ saturates quickly to $7$ for $\tau >
1$ , and to $0$ for $\tau < 1$ \cite{csaki}.
\section{\label{colli}Radion production in $\gamma e^-$ collisions }

 High energy $\gamma e^-$ colliders  have been
essential instruments to search for the fundamental constituents
of matter and their interactions. The source of high energy photon
provided in experiments is by using the laser backscattering
technique \cite{gin,gin1}. In our earlier work \cite{jhep}, we
have considered the radion production in external electromagnetic
fields. This section is devoted to
 the production of radions in the high energy $\gamma e^-$
 collisions. We now consider the process in which the initial
 state contains an electron and a photon and the final state
 contains a pair of electron and radion,
 \be   e^-(p_1,\lambda') + \gamma (p_2,\lambda )
\rightarrow \ e^-(k_1,\tau ) \ + \phi(k_2). \label{p1}
\label{pro1} \ee

Here $p_i$, $k_i$($i=1,2$) stand for the momentum and
$\lambda$,$\lambda'$, $\tau$ are the  helicity of the particle,
respectively. There are three Feynman diagrams contributing to
reaction ($\ref{p1}$) , representing the {\it s, u, t} - channel
exchange depicted in Fig.1.

\begin{center}
\begin{picture}(400,50)(0,0)
\Photon(-10,35)(5,10){2}{6} \ArrowLine(-10,-15)(5,10)
\ArrowLine(5,10)(55,10) \DashLine(55,10)(70,35){2}
\ArrowLine(55,10)(70,-15) \Text(32,18)[]{$e^{-} $}
\Text(-16,40)[]{$\gamma$} \Text(-16,-20)[]{$e^-$}
\Text(80,40)[]{$\phi$} \Text(80,-20)[]{$e^-$} \Text(185,-42)[]{
Fig.1: Feynman diagrams for $\ e^- \gamma \rightarrow \phi \ e^-
$}

%u-channel

\Photon(150,35)(200,10){2}{12} \DashLine(150,10)(200,35){2}
\ArrowLine(135,-15)(150,10) \ArrowLine(150,10)(200,10)
\ArrowLine(200,10)(215,-15)

\Text(180,5)[]{$e^{-} $} \Text(145,40)[]{$\gamma$}
\Text(130,-20)[]{$e^-$} \Text(220,40)[]{$\phi$}
\Text(215,-20)[]{$e^-$}
%t-channel
\Photon(270,35)(310,35){2}{6} \Photon(310,-15)(310,35){2}{8}
\DashLine(350,35)(310,35){2} \ArrowLine(270,-15)(310,-15)
\ArrowLine(310,-15)(350,-15)

\Text(330,10)[]{$\gamma$} \Text(265,40)[]{$\gamma$}
\Text(265,-20)[]{$e^-$} \Text(365,40)[]{$\phi$}
\Text(360,-20)[]{$e^-$}
\end{picture}
\end{center}
%%%%%%%%%%%%%%%%%%%%%%%%%
\vspace*{1.5cm} The amplitude for this process can be written as
\begin{equation}
M_{i}=\epsilon_{\mu}(p_2)\overline{u}(k_1)A^{\mu}_i u(p_1), i=
s,u, t
\end{equation}
where $\epsilon_{\mu}(p_2)$ are the polarization vector of the
$\gamma$ photon. In the high energy limit $s >> m_e^2$, assuming a
vanishing mass of electron, the $A^{\mu}_i$ for the three diagrams
are given by
\begin{equation}
A^{\mu}_{s}=\frac{-ie m_{e}}{\La_\phi
q^2_{s}}\not{q_s}\gamma^{\mu},
\end{equation}
%$q_1=p_1+p_2=k_1+k_2$.where $q_1=p_1+p_2=k_1+k_2$. And $u$-channel is
\begin{equation}
A^{\mu}_{u}=\frac{-ie m_{e}}{\La_\phi
q^2_{u}}\gamma^{\mu}\not{q_u},
\end{equation}
\begin{equation}
A^{\mu}_{t} = \frac{4 e
c_{\phi\gamma\gamma}}{q^2_{t}}[(p_2.q_t)\gamma^{\mu}
-\not{p_2}q_{t}^{\mu}].
\end{equation}
Here,  $q_s=p_1+p_2=k_1+k_2$, $q_u=p_1-k_1=k_2-p_2$,
$q_t=p_1-k_2=k_1-p_2$, and $s = (p_1 + p_2)^2$ is the square of
the collision energy. We  work in the center-of-mass frame and
denote the scattering angle by  $\theta $ (the angle between
momenta of the initial and the final electrons).
We give some estimates of the cross - section as follows\\

 \hs i) We show in Fig.2 the behavior of $d\sigma$/$\cos \theta $
 at the fixed collision energy,  $\sqrt{s}= 3$ TeV (CLIC). We have chosen
 a relatively low value of the radion mass, $m_{\phi} = 10$ GeV
 and the polarization coefficients are $P_{e} = 0$, $0.5$ and $1$,
respectively. From Fig.2 we see that $d\sigma$/$\cos \theta$ is
peaked in the forward direction ($\theta \approx 0$)
% (this is due to the $e^-$ poleterm in the $u$ - channel)
 but it is flat in the backward direction.\\
\hs ii) In Fig.3 we plot cross section as a function of the
collision energy and the polarized coefficient of electron. The
figure shows that the cross section has the maximum value
($\sigma_{max}$) at $P_{e}=\pm 1$ and has the minimum value
($\sigma_{0}$) at $P_{e}= 0$, it is worth noting that
$\sigma_{max} = 2\sigma_{0}$. At CLIC we get $\sigma_{max}=
2.2743\times10^{-3}$ pb for $m_{\phi}= 10$ GeV, which is  smaller
than production cross section of the bilepton and $Z'$ in the
3-3-1 models~\cite{stl,sblk}, but it is large enough to measure
the radion production.\\
 \hs iii) The radion mass dependence of
the cross-section $\sigma$ at fixed energy as in i), $\sqrt{s} =
3$ TeV, are shown in Fig.4. The polarized coefficient is chosen as
$P_{e}=1$ and the mass range is $10$ GeV $\leq m_{\phi}\leq 500$
GeV. With the high integrated luminosity $L = 9\times 10^4
fb^{-1}$~\cite{stl}, the number of events with some different
values of the radion mass is given in Table 1. From these results,
we can see that with the high integrated luminosity and at the
high degree of polarization, the production
cross section of the radions may give observable values at CLIC.\\

\begin{table}[h]
\bc
\begin{tabular}{|c|c|c|c|c|c|c|}
  \hline
 $m[\mathrm{GeV}]$ & 10 & 100& 200& 300& 400& 500 \\
  \hline
  $N$ &$312490$& $ 312169$
 & $ 311198$ &$ 309587$ & $307347$& $304494$\\
 \hline
\end{tabular}
\caption{\label{tab1}The number of events with some different
values of the radion mass. } \ec
\end{table}

\section{\label{culc}Conclusion}
\hspace*{0.5cm}In this work, we have evaluated the radion
production in $\gamma e^- $ collisions. The result shows that with
the high integrated luminosity and the high polarization of
electron beams, the production cross section of radions may give
observable values at the moderately high energies in future
colliders. Note that here we focus only on the case of the radion
mass in the range of $GeV$, which is large enough to avoid radion-
mediated flavor changing neutral currents~\cite{aga}. In this mass
range, the radion signal at $\gamma e^- $ colliders can be
compared with the high energy colliders at CERN LHC~\cite{gon}, in
which the detection of radions will be the first signature
of the RS model.\\
\textbf{Acknowledgements:} The work is supported in part by the
National Foundation for Science and Technology Development
(NAFOSTED) of Vietnam under Grant No. $103.03-2012.80$.\\

\newpage
\begin{figure}\bc
\includegraphics[width=10cm]{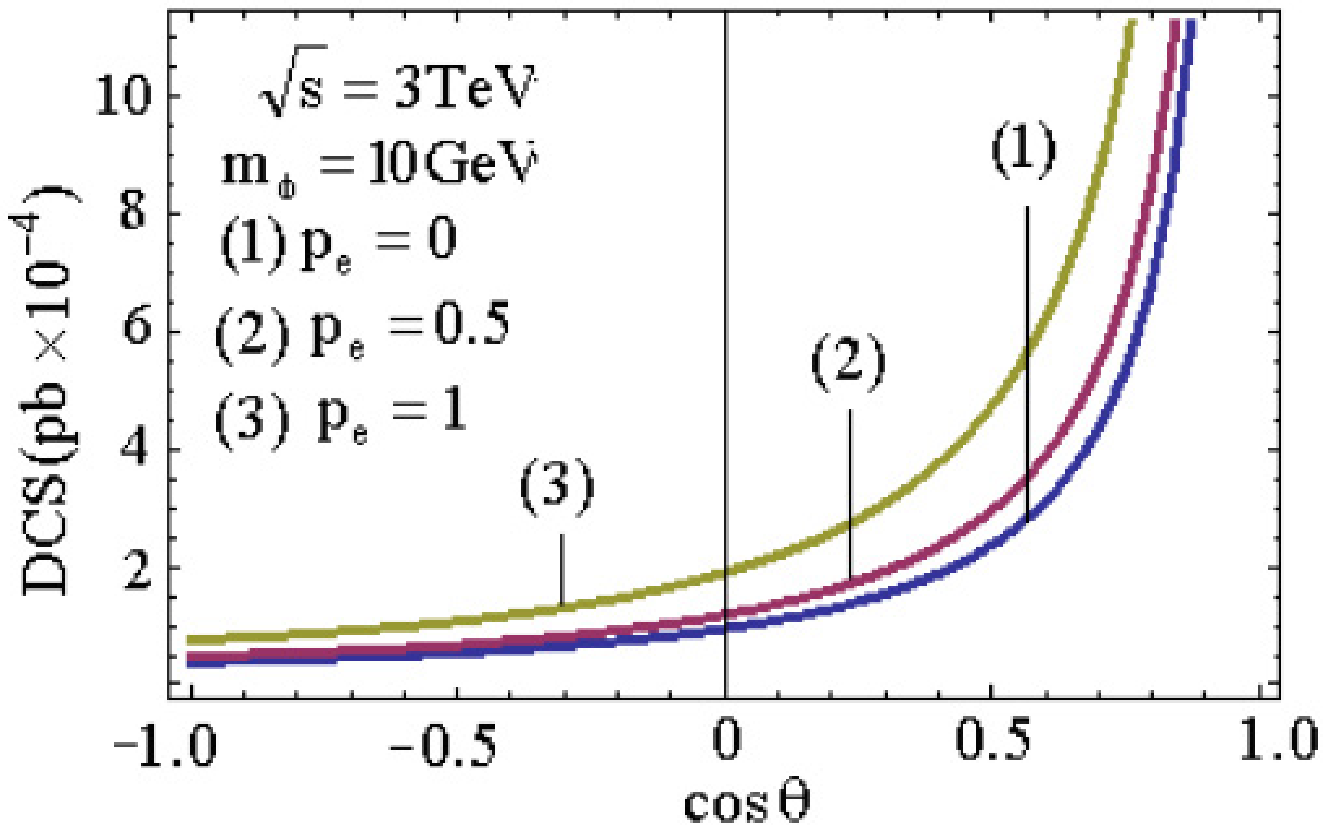}\\
{Fig.2: Different cross-section of the process $\gamma
e^-\rightarrow \phi e^-$ as a function of $\cos\theta$ at
$\sqrt{s}= 3$ TeV. The radion mass is taken to be $100$ GeV and
the polarization coefficients are chosen as $P_{e} = 0$, $0.5$ and
$1$, respectively.} \ec
\end{figure}
\begin{figure}\bc
\includegraphics[width=10cm]{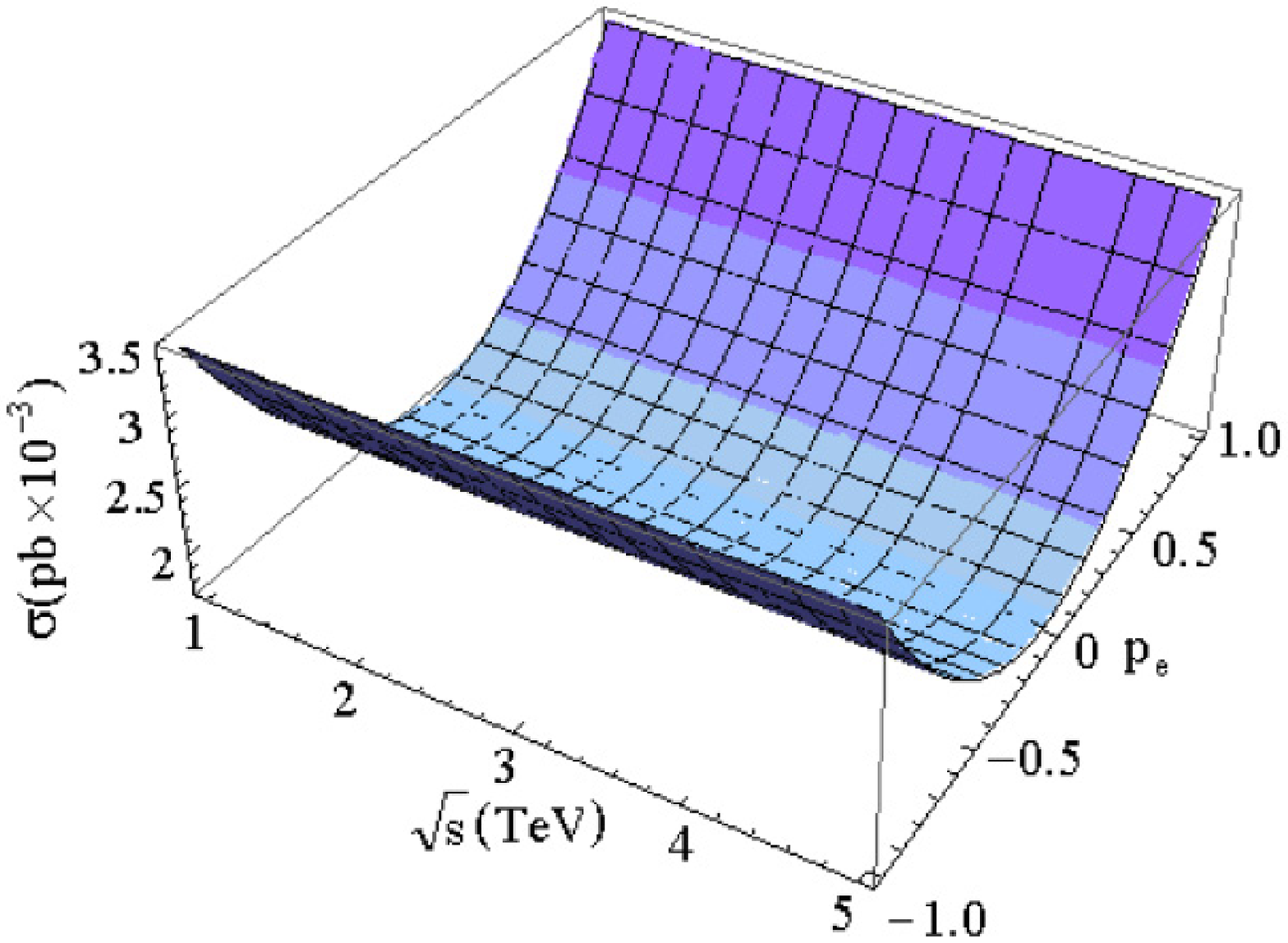}\\
{Fig.3: Cross-section of the process $\gamma e^- \rightarrow \phi
e^- $ as a function of the polarization coefficient $P_{e}$  and
the collision energy $\sqrt{s}$.}\ec
\end{figure}
\begin{figure}\bc
\includegraphics[width=10cm]{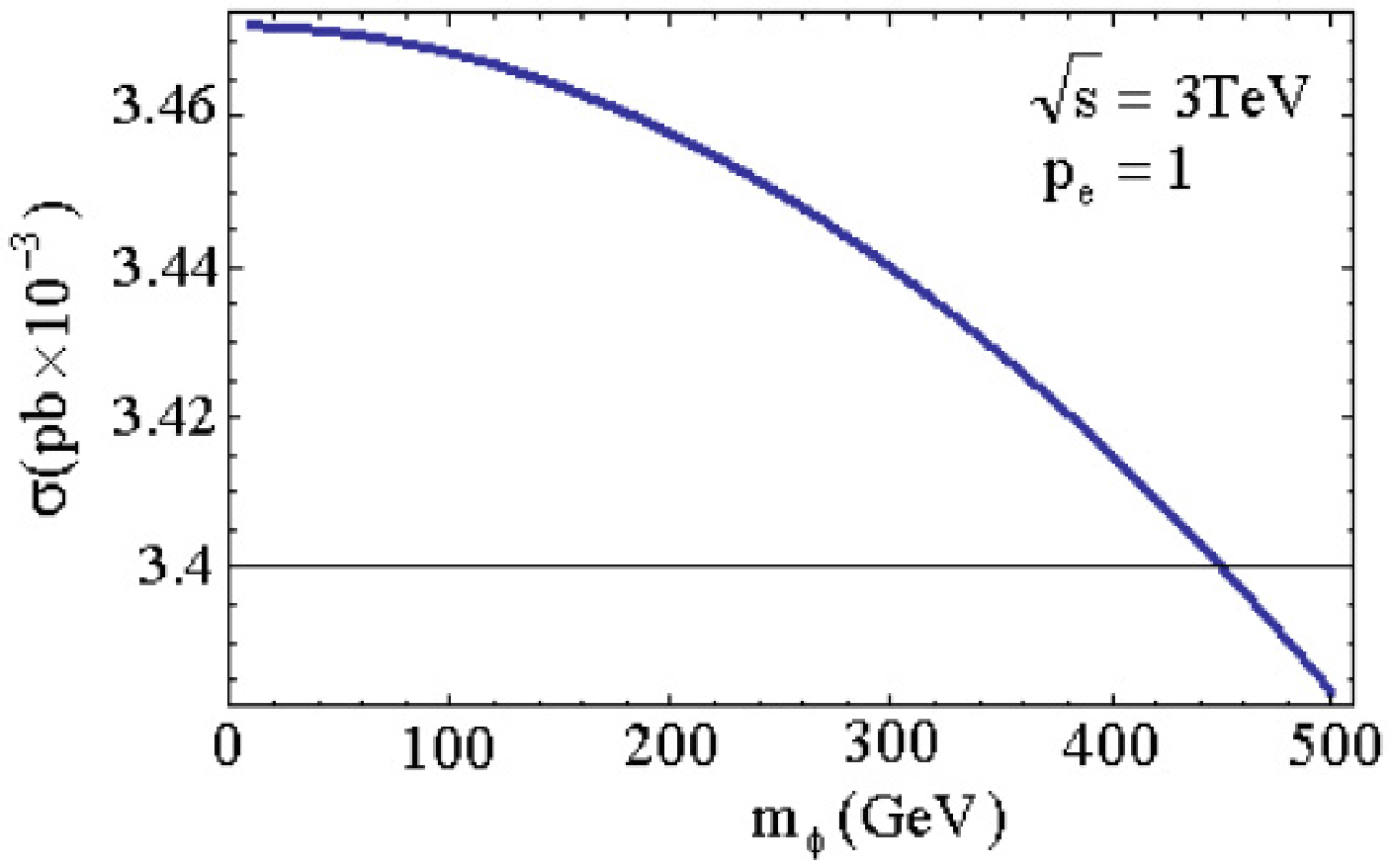}\\
{Fig.4: Cross-section of the process $\gamma e^- \rightarrow \phi
e^- $ as a function of the radion mass at $\sqrt{s}= 3$ TeV. The
polarization coefficient is chosen as $P_{e} = 1$.}\ec
\end{figure}

\end{document}